\documentclass[pra,amsfonts, twocolumn]{revtex4}
\bibpunct{[}{]}{,}{a}{}{;}
\RequirePackage{times}
\RequirePackage{mathptm}
\begin{document}
\title{Time generated by intrinsic observers\footnote{published in ``Cybernetics and Systems 96, Proceedings of the 13th European Meeting on Cybernetics and Systems Research, ed. by Robert Trappl (Austrian Society for Cybernetic Studies, Vienna, 1996), pp. 162-166''}}
\author{Karl Svozil}
\email{svozil@tuwien.ac.at}
\homepage{http://tph.tuwien.ac.at/~svozil}
\affiliation{Institute for Theoretical Physics, Vienna University of Technology,  \\
Wiedner Hauptstra\ss e 8-10/136, A-1040 Vienna, Austria}
\begin{abstract}
 We shortly review the construction of knowledge by intrinsic observers. Intrinsic observers are embedded in a system and are inseparable parts
thereof.  The intrinsic viewpoint has to be contrasted with an extrinsic, "God's eye" viewpoint, from which the system can be observed externally without in any way changing it.  This epistemological distinction has concrete, formalizable consequences.  One consequence is the emergence of  "complementarity" for intrinsic observers, even if the underlying system is totally deterministic (computable).  Another consequence is the appearence of time and inertial frames for intrinsic observers.  The necessary operational techniques are developed in the context of  Cellular Automata.  We finish with a somewhat speculative question.  Given space-time frames generated by clocks which use sound waves for synchronization; why could supersonic travel not cause time paradoxes?
\end{abstract}

\maketitle

\section{Operational concepts and intrinsic observers}

John Casti posed the following scenario
\cite{casti-priv}
(for a
related discussion, see
\cite{sv:86a}),
{\em
``[[Imagine some creatures---]]they may be carbon-based creatures just
like you and me. The difference is that they live in a world in which the
primary sensory inputs are not from the electromagnetic spectrum like
light, but rather come from sound waves. Note that this is not simply a
world of the blind; rather, it is a world in which there are no sensory
organs for perceiving any part of the electromagnetic spectrum. In this
case, then, $\ldots$ such creatures would see the speed of sound
as a fundamental barrier to the velocity of any material object. Yet we
as creatures that {\em do} possess sensory organs for perceiving the
electromagnetic
spectrum see the sound barrier as no fundamental barrier at all. So, by
analogical extension, there may be creatures ``out there'' who regard the
speed of light as no more of a barrier than we regard the speed of
sound.''
}

Casti's scenario is one in which
the observers are embedded
in a universe which, to them, solely consists of sound waves. Such
intrinsic observers develop a description
\cite{bos,toffoli:79,roessler1,sv:83,sv:86,sv:86a,sv:93}
of the
universe which may appear drastically different (cf. the
emergence of complementarity
\cite[chapter 10]{sv:93}) from what some hypothetical
``super-observer''
perceives, who is peeking at the system from a ``God's eye,'' extrinsic
position. One of the first researchers
bothering about these issues has been the 18th century physicist
Boskovich,
{\em ``$\ldots$ And we would have the same impressions if, under
conservation of distances, all directions would be rotated by the same
angle, $\ldots$ And even if the distances themselves would be
decreased, whereby the angles and the proportions would be conserved,
$\ldots$: even then we
[[the
observers]]
would have no changes in our
impressions.
$\ldots$ A movement, which is common to us [[the observers]] and
to the Universe, cannot be observed by us; not even if everything would
be stretched or shrinked by an arbitrary amount.''
}

Although generally not presented that way,
relativity theory consists of two distinct parts: (i) it deals with
conventions about how to operationalize certain concepts such as ``equal
time at spatially separated points,'' in particularly synchronization
procedures; and
(ii) it states physical assumptions about the invariance of certain
phenomena, such as the speed of light, and the
laws of theoretical physics under changes of reference frames.
Whereas the former conventions are mere working definitions,
the latter statements are supposed to be God-given, eternal
symmetries. Conventions can be changed at the price of complicating the
theoretical formalism by means of a non-optimal representation of
theory.
Phenomena are a matter of physical fact.
For instance,
a preferred frame of reference, e.g., the one at rest with
respect to cosmic background radiation, could be artificially
introduced.
To put it in the words of the late John Bell (cf. \cite{bell:92}, p.
34):
{\em ``You can pretend that whatever inertial frame you have chosen is
the ether of the 19th century physicists, and in that frame you can
confidently apply the idea of the FitzGerald contraction [[of length of
material bodies such as scales]],  Larmor [[time]] dilation and
Lorentz lag. It is a great pity that students dont understand this.
$\ldots$''}


It was the radical operational feature in Einstein's theory of special
relativity, which stimulated the physicist Bridgman.
 Bridgman demanded that the meaning
of theoretical concepts should ultimately be based upon entities which
are intrinsically representable and operational. That is,
(cf. \cite{bridgman-reflextions}, p. V),
{\em ``the meaning of one's terms are to be found
by an analysis of the operations which one performs in applying the
term in concrete situations or in veryfying the truth of statements or
in finding the answers to questions.''}
More specifically (cf. \cite{bridgman}, p. 103),
{\em ``$\ldots$ the
meaning of length is to be sought in those operations by which the
length of physical objects is determined, and the meaning of
simultaneity is sought in those physical operations by which it is
determined whether two physical events are simultaneous or not.''}

Therefore, we are free to choose whatever physical concepts seem
appropriate as long as they are operational. In particular, Casti's
sound-perceiving creatures might choose sound, the author's water-fleas
may
use water waves, and human physicists may use light for the purpose of
Einstein synchronization. The resulting frame of references will be
different in all these cases. The space-time parameters by which events
and phenomena are represented will be different, too.
Using different types of formal representations
of the same physical
system---one may speak of
layers or levels of description---might be an effective way to
grasp different features
of that system (associated, for instance, with different levels of
complexity). This resembles Anderson's thesis of
``emerging laws'' (\cite{anderson}, p. 193; cf. Schweber
\cite{schweber}),
{\em ``The ability to reduce everything to simple fundamental laws does
not imply the ability from these laws and reconstruct the universe.
$\ldots$ The constructionist hypothesis breaks down when confronted
with the twin difficulties of scale and complexity. $\ldots$ at each
level of complexity, entirely new properties appear, and the
understanding of the new behaviors requires research which I think is
as fundamental in nature as any other.''}
It is to be expected that in such an organization of physical concepts,
the notion of causality need not be consistently defined for all
descriptions {\em combined:} what may appear as causal connection
(i.e., cause
and effect) in one description needs not be causally connected in
another description. Therefore it is of great relevance to make precise
the limits and applicability of each of these descriptions.

\section{Space-time frames in Cellular Automata}
In what follows, an explicit example for the
construction of space-time frames in com\-pu\-ter-\-ge\-nera\-ted
universes
which are generated by {\em intrinsic} procedures and observations will
be given.
The onedimensional Cellular Automaton
(CA) models
\cite{von-neumann:burks,fredkin}
considered here are equivalent to any other universal
computing agent but have the advantage of good representability on the
twodimensional printing page combined with easy programmability.

Our primary concern will be the explicit construction of
intrinsic space-time frames
by adopting Einstein's synchronization conventions.
Stated pointedly, we are interested primarily with ``virtual reality''
physics and physical epistemology.
We do {\em not} attempt to reconstruct relativity theory
one-to-one  in the cellular automaton context.
In particular, no
Lorentz-invariant kinematic theory is introduced.
Therefore, certain physical statements, in particular the relativity
principle, stating that all laws of physics have an identical
form in all inertial frames, needs not to be satisfied
(cf.
\cite{bia}).

However, it has been argued for quite some time \cite{fredkin}
that, since all
``construable''
(in the sense of
``recursively enumerable'')
universes are realisable within the CA framework, also the
special theory of relativity can be implemented on such a structure.

%
%
%
%

\subsection{Synchronization}

 Without loss of generality it is assumed that the maximal velocity,
 denoted by $c$, by which a body of information can move is one
 cell per cycle time.
 Flows of this kind will be called {\em rays}.
 In analogy to relativity theory, this velocity can be used to define
 clocks or
 synchronized events. We shall start by an explicit model of a ray
 clock. It consists of two mirrors, denoted by {\tt I}, and a ray of
velocity 1 cell per time cycle, denoted by $>$ and $<$, which is
constantly reflected back and forth between the two mirrors, and a
one-place digital display right to the right mirror. In this model,
after each
backreflection of the light ray, the digit on the display increases by
one modulus 10.
The explicit transformation rules for a onedimensional CA with these
properties are listed in appendix A.
The time evolution of a ray clock is drawn in Fig. \ref{ray-clock}.
\begin{figure}
 {\tiny
 \begin{verbatim}
 ... __I>__I0__ ...          ... __I>__I4__ ...         ... __I>__I8__ ...
 ... __I_>_I0__ ...          ... __I_>_I4__ ...         ... __I_>_I8__ ...
 ... __I__>I0__ ...          ... __I__>I4__ ...         ... __I__>I8__ ...
 ... __I__<*0__ ...          ... __I__<*4__ ...         ... __I__<*8__ ...
 ... __I_<_I1__ ...          ... __I_<_I5__ ...         ... __I_<_I9__ ...
 ... __I<__I1__ ...          ... __I<__I5__ ...         ... __I<__I9__ ...
 ... __I>__I1__ ...          ... __I>__I5__ ...         ... __I>__I9__ ...
 ... __I_>_I1__ ...          ... __I_>_I5__ ...         ... __I_>_I9__ ...
 ... __I__>I1__ ...          ... __I__>I5__ ...         ... __I__>I9__ ...
 ... __I__<*1__ ...          ... __I__<*5__ ...         ... __I__<*9__ ...
 ... __I_<_I2__ ...          ... __I_<_I6__ ...         ... __I_<_I0__ ...
 ... __I<__I2__ ...          ... __I<__I6__ ...         ... __I<__I0__ ...
 ... __I>__I2__ ...          ... __I>__I6__ ...         ... __I>__I0__ ...
 ... __I_>_I2__ ...          ... __I_>_I6__ ...         ... __I_>_I0__ ...
 ... __I__>I2__ ...          ... __I__>I6__ ...         ... __I__>I0__ ...
 ... __I__<*2__ ...          ... __I__<*6__ ...         ... __I__<*0__ ...
 ... __I_<_I3__ ...          ... __I_<_I7__ ...         ... __I_<_I1__ ...
 ... __I<__I3__ ...          ... __I<__I7__ ...         ... __I<__I1__ ...
 ... __I>__I3__ ...          ... __I>__I7__ ...               .
 ... __I_>_I3__ ...          ... __I_>_I7__ ...               .
 ... __I__>I3__ ...          ... __I__>I7__ ...               .
 ... __I__<*3__ ...          ... __I__<*7__ ...
 ... __I_<_I4__ ...          ... __I_<_I8__ ...
 ... __I<__I4__ ...          ... __I<__I8__ ...
 \end{verbatim}  }
\caption{Ray clock. \label{ray-clock}}
\end{figure}

A very similar configuration as for the light clock can be used as a
device for Einstein synchronization \cite{einstein:1905,sv:93}:
 Assume two clocks at two arbitrary points $A$ and $B$ which are ``of
similar
 kind.''
 At some arbitrary $A$-time $t_A$ a  ray goes from $A$ to $B$. At $B$
it is instantly (without delay)
 reflected  at $B$-time $t_B$ and reaches $A$ again at
 $A$-time $t_{A'}$. The clocks in $A$ and $B$ are {\em synchronized}
 if
\begin{equation}
 t_B-t_A=t_{A'}-t_B \quad .
\label{e:1}
\end{equation}
 The two-ways ray velocity is given by
\begin{equation}
 {2 \vert {AB}\vert \over t_{A'}-t_A}=c \quad ,
\end{equation}
 where $\vert AB\vert $ is the distance between $A$ and $B$.

 The ray velocity can then be
{\em defined} to be identical for all frames, irrespective of
whether
 they are moving with respect to the rest frame of the cellular
 space or not.
Of course, this invariance of the ray speed with respect to
changes of coordinate systems should ultimately be motivated by
phenomenology and a proper choice of conventions. E.g., in relativity
theory, the invariance of the Maxwell
equations with respect to conformal, or angle preserving, coordinate
transformations in four dimensions assures
that for light-like vectors, $ds^2=dx^2-(c\,dt)^2=0$ and thus $x=ct$.

For synchronization, the same CA transformation rules as for the ray
clock, which are listed in appendix
\ref{a:lightclock} can be used.
In Fig. \ref{synchro}, an example of synchronization between two clocks
$A$ and $B$
is drawn.
\begin{figure}
 {\tiny
 \begin{verbatim}
      clock1  A       B  clock2

... __I>__I0__I>______I__I>__I0__ ...
... __I_>_I0__I_>_____I__I_>_I0__ ...
... __I__>I0__I__>____I__I__>I0__ ...
... __I__<*0__I___>___I__I__<*0__ ...
... __I_<_I1__I____>__I__I_<_I1__ ...
... __I<__I1__I_____>_I__I<__I1__ ...
... __I>__I1__I______>I__I>__I1__ ...
... __I_>_I1__I______<*__I_>_I1__ ...
... __I__>I1__I_____<_I__I__>I1__ ...
... __I__<*1__I____<__I__I__<*1__ ...
... __I_<_I2__I___<___I__I_<_I2__ ...
... __I<__I2__I__<____I__I<__I2__ ...
... __I>__I2__I_<_____I__I>__I2__ ...
... __I_>_I2__I<______I__I_>_I2__ ...
... __I__>I2__I>______I__I__>I2__ ...
... __I__<*2__I_>_____I__I__<*2__ ...
... __I_<_I3__I__>____I__I_<_I3__ ...
... __I<__I3__I___>___I__I<__I3__ ...
... __I>__I3__I____>__I__I>__I3__ ...
... __I_>_I3__I_____>_I__I_>_I3__ ...
... __I__>I3__I______>I__I__>I3__ ...
    .
    .
    .
 \end{verbatim}  }
\caption{Synchronization by ray exchange. \label{synchro}}
\end{figure}

\subsection{Moving reference frames}

This section deals with what happens with the intrinsic
synchronization and the space-time
coordinates when observers are considered which move with respect to the
CA medium. For simplicity, assume constant motion of $v$ automaton cells
per time cycle. With these units, the ray speed is $c=1$, and $v\le 1$.

There are numerous ways to simulate sub-ray motion on a CA. In what
follows, the case $v=1/3$ will be studied in such a way that every three
CA time cycles the walls, symbolised by ${\tt I}$, move one cell to the
right. [Strictly speaking, there should be a periodic transformation of
the wall such that ${\tt I}\rightarrow a\rightarrow b\rightarrow {\tt
I}$ and the states $a$ and $b$ have the same reflection properties as
${\tt I}$.]

Notice that two clocks which
are synchronized in a reference frame which is at rest with respect to
the CA medium are {\em not synchronized} in their own co-moving
reference frame.
Consider, as an example, the CA drawn in Fig.
\ref{f:3}(a). (Strictly speaking, the CA rule here depends on
a two-neighbor interaction.)
  By evaluating equation
(\ref{e:1}) for $t_A=1$, $t_B=4$, $t_{A'}=5$, and $4-1\neq 5-4$.
If the first clock is corrected to make up for the different time of ray
flights as in Fig. {f:3}(b),
$t_A=2$, $t_B=4$, $t_{A'}=6$, and $4-2 = 6-4$.
This correction, however, is the reason for asynchronicity of the two
ray clocks with respect to the ``original'' CA medium.

\begin{figure}[t]
 {\tiny
 \begin{verbatim}
__I>_I0__I___<___I__I>_I0________________   __I>_I1__I___<___I__I>_I0________________
__I_>I0__I__<____I__I_>I0________________   __I_>I1__I__<____I__I_>I0________________
__I_<*0__I_<_____I__I_<*0________________   __I_<*1__I_<_____I__I_<*0________________
___I>_I1__I>______I__I>_I1_______________   ___I>_I2__I>______I__I>_I1_______________
___I_>I1__I_>_____I__I_>I1_______________   ___I_>I2__I_>_____I__I_>I1_______________
___I_<*1__I__>____I__I_<*1_______________   ___I_<*2__I__>____I__I_<*1_______________
____I>_I2__I__>____I__I>_I2______________   ____I>_I3__I__>____I__I>_I2______________
____I_>I2__I___>___I__I_>I2______________   ____I_>I3__I___>___I__I_>I2______________
____I_<*2__I____>__I__I_<*2______________   ____I_<*3__I____>__I__I_<*2______________
_____I>_I3__I____>__I__I>_I3_____________   _____I>_I4__I____>__I__I>_I3_____________
_____I_>I3__I_____>_I__I_>I3_____________   _____I_>I4__I_____>_I__I_>I3_____________
_____I_<*3__I______>I__I_<*3_____________   _____I_<*4__I______>I__I_<*3_____________
______I>_I4__I______>I__I>_I4____________   ______I>_I5__I______>I__I>_I4____________
______I_>I4__I______<*__I_>I4____________   ______I_>I5__I______<*__I_>I4____________
______I_<*4__I_____<_I__I_<*4____________   ______I_<*5__I_____<_I__I_<*4____________
_______I>_I5__I___<___I__I>_I5___________   _______I>_I6__I___<___I__I>_I5___________
_______I_>I5__I__<____I__I_>I5___________   _______I_>I6__I__<____I__I_>I5___________
_______I_<*5__I_<_____I__I_<*5___________   _______I_<*6__I_<_____I__I_<*5___________
________I>_I6__I>______I__I>_I6__________   ________I>_I7__I>______I__I>_I6__________
________I_>I6__I_>_____I__I_>I6__________   ________I_>I7__I_>_____I__I_>I6__________
________I_<*6__I__>____I__I_<*6__________   ________I_<*7__I__>____I__I_<*6__________
_________I>_I7__I__>____I__I>_I7_________   _________I>_I8__I__>____I__I>_I7_________
_________I_>I7__I___>___I__I_>I7_________   _________I_>I8__I___>___I__I_>I7_________
_________I_<*7__I____>__I__I_<*7_________   _________I_<*8__I____>__I__I_<*7_________
__________I>_I8__I____>__I__I>_I8________   __________I>_I9__I____>__I__I>_I8________
__________I_>I8__I_____>_I__I_>I8________   __________I_>I9__I_____>_I__I_>I8________
__________I_<*8__I______>I__I_<*8________   __________I_<*9__I______>I__I_<*8________
___________I>_I9__I______>I__I>_I9_______   ___________I>_I0__I______>I__I>_I9_______
___________I_>I9__I______<*__I_>I9_______   ___________I_>I0__I______<*__I_>I9_______
___________I_<*9__I_____<_I__I_<*9_______   ___________I_<*0__I_____<_I__I_<*9_______
____________I>_I0__I___<___I__I>_I0______   ____________I>_I1__I___<___I__I>_I0______
____________I_>I0__I__<____I__I_>I0______   ____________I_>I1__I__<____I__I_>I0______
____________I_<*0__I_<_____I__I_<*0______   ____________I_<*1__I_<_____I__I_<*0______
_____________I>_I1__I>______I__I>_I1_____   _____________I>_I2__I>______I__I>_I1_____
_____________I_>I1__I_>_____I__I_>I1_____   _____________I_>I2__I_>_____I__I_>I1_____
_____________I_<*1__I__>____I__I_<*1_____   _____________I_<*2__I__>____I__I_<*1_____
______________I>_I2__I__>____I__I>_I2____   ______________I>_I3__I__>____I__I>_I2____
______________I_>I2__I___>___I__I_>I2____   ______________I_>I3__I___>___I__I_>I2____
______________I_<*2__I____>__I__I_<*2____   ______________I_<*3__I____>__I__I_<*2____
_______________I>_I3__I____>__I__I>_I3___   _______________I>_I4__I____>__I__I>_I3___
_______________I_>I3__I_____>_I__I_>I3___   _______________I_>I4__I_____>_I__I_>I3___
_______________I_<*3__I______>I__I_<*3___   _______________I_<*4__I______>I__I_<*3___
________________I>_I4__I______>I__I>_I4__   ________________I>_I5__I______>I__I>_I4__
________________I_>I4__I______<*__I_>I4__   ________________I_>I5__I______<*__I_>I4__
________________I_<*4__I_____<_I__I_<*4__   ________________I_<*5__I_____<_I__I_<*4__
          .                                           .
          .                                           .
          .                                           .
(a)                                         (b)
 \end{verbatim}  }
\caption{Unsynchronized (a) and synchronized (b) ray clocks in moving
reference frame. \label{f:3}}
\end{figure}

Let us now explicitly construct the coordinate axes ${\overline x}$ and
${\overline t}$ of a system which moves with constant velocity with
respect to the medium.
For convenience, let us switch to continuous coordinates. That is, the
``grainyness'' of the CA medium
is disregarded; i.e., coordinates will be represented by real numbers.
In this construction, the convention of the constancy
of the speed
of rays for all reference frames will play an important role. The new
space
and time axes will both become rotated towards the ray coordinates.

Consider Fig. \ref{f:4}, which represents the process drawn in
Fig.
\ref{f:3}(b).
\begin{figure}
\unitlength 0.50mm
\special{em:linewidth 0.4pt}
\linethickness{0.4pt}
\begin{picture}(80.00,80.00)
\put(10.00,10.00){\vector(0,1){70.00}}
\put(10.00,10.00){\vector(1,0){70.00}}
\put(10.00,10.00){\line(1,1){70.00}}
\put(10.00,10.00){\vector(1,2){35.00}}
\put(10.00,10.00){\vector(2,1){70.00}}
\put(30.00,10.00){\line(1,2){35.00}}
\put(50.00,50.00){\line(-5,6){12.33}}
\put(23.33,36.67){\circle*{1.49}}
\put(37.33,65.33){\circle*{1.33}}
\put(50.00,50.00){\circle*{1.49}}
\put(10.00,30.00){\line(2,1){70.00}}
\put(5.67,75.00){\makebox(0,0)[cc]{$t$}}
\put(38.00,75.33){\makebox(0,0)[cc]{${\overline t}$}}
\put(76.00,37.67){\makebox(0,0)[cc]{${\overline x}$}}
\put(76.00,4.33){\makebox(0,0)[cc]{$x$}}
\put(79.66,76.67){\makebox(0,0)[lc]{$x-t={\overline x}-{\overline t}=0$}}
\put(10.00,5.00){\makebox(0,0)[cc]{$0$}}
\put(10.00,10.00){\circle*{1.49}}
\put(6.33,10.00){\makebox(0,0)[cc]{$A$}}
\put(53.00,46.33){\makebox(0,0)[cc]{$B$}}
\put(32.00,65.33){\makebox(0,0)[cc]{$A'$}}
\put(20.33,40.67){\makebox(0,0)[cc]{$1$}}
\put(30.00,4.67){\makebox(0,0)[cc]{$a$}}
\end{picture}
\caption{Construction of the coordinate axes ${\overline x}$ and
${\overline t}$ of a system which moves with constant velocity with
respect to the medium.
\label{f:4}}
\end{figure}
Assume that the two mirrors are $a$ (arbitrary) units apart.
For the system at rest with respect to the CA medium, a ray is emitted
from the origin
$A=(0,0)$
and arrives at the mirror at $B=(vt_B+a,t_B)$, where it is reflected
and arrives at the original mirror at $A'=(vt_{A'}, t_{A'})$.
Since $B$ lies on the ray which comes from the origin,
$B=(t_B,t_B)=\left({a\over 1-v},{a\over 1-v}\right)$.
Since $A'$ lies on the ray which comes from $B$,
$t_{A'}=-x_{A'}+C$ and
$t_{B}=-x_{B}+C$. By evaluating $C$, one obtains
$A'=\left({2va\over 1-v^2},{2a\over 1-v^2}\right)$.
In the coordinate system which is moving with respect to the
CA medium, ${\overline A}=(0,0)$. In order for the clocks to be
synchronized, i.e., by equation
(\ref{e:1}),
 ${\overline t_B} =({\overline
t_{A'}}+{\overline t_A})/2$.
But this should also be the time coordinate ${\overline t_1}$, since
this is just half the time from $A$ to $A'$. Thus one obtains two points
(events)
$1=\left({vt_{A'}\over 2},{t_{A'}\over 2}\right)$ and $B$ whose
time coordinates in
the moving reference frame is identical; i.e., ${\overline
t_1}={\overline t_B}$.
 A short calculation shows that, with respect to
the coordinate system which is at rest in the CA
medium, the
lines of equal time coordinates for a system wich is moving with
constant velocity $v$, e.g., the
${\overline x}$-axis, have slope $1/v$, whereas the
lines of equal space coordinates, e.g., the ${\overline t}$-axis, have
slope $v$.

These transformation of coordinate axes correspond to the
Lorentz transformations
\begin{equation}
{\overline t} =
{t-vx\over \sqrt{1-v^2}}\quad , \quad
{\overline x} =
{x-vt\over \sqrt{1-v^2}}\quad .\label{el:rt2}
\end{equation}
The specific form of the transformation (\ref{el:rt2})
comes as no surprise, since it has been derived by implicitly assuming
that constant
motion transforms into constant motion and that the ray speed is the
same for all reference frames; both conditions being the kinematic
equivalent to the relativity principle.

\subsection{Sub-ray synchronization}

So far, only rays propagating one CA cell per cycle have been
considered.
It is not entirely unreasonable to assume
synchronization with signals which propagate slower (or faster) than
these rays.

A
typical example would be the use of a signal for synchronization with
slower-than-ray speed, i.e., $c'<c$.

One consequence of sub-ray synchronization is the possibility of
``super-ray'' speeds $v$ such that
$c\ge v>c'$, and of a
``time
travel'' with respect to such space-time frames. That is, for certain
observers moving with
$v>c'$, the time coordinate defined by $c'$ would ``run backward.''
This ``time travel'' however, is nothing particularly mysterious, but
the
outcome of the specific synchronization convention chosen (cf. below).

\section{Can supersonic travel give rise to time paradoxes?}

Let us come back to Casti's sound-sensitive creatures---suppose that
they
call themselves {\sc smorons}. Suppose further that they discover the
possibility to generate and detect light; e.g., by sonoluminescence.
 It can be expected that this discovery will cause a major
trauma for the {\sc smorons}, because they will find out that they could
communicate much faster than by sound waves, on which their space-time
frames are based.
If they have applied the Einstein conventions for defining space-time
frames, they would observe light as a supersonic (i.e.,
faster-than-sound) signal, which propagates on space-like
($(\Delta x)^2-(\Delta t)^2>0$) world lines. Such a phenomenon which
might
allow forward in time signalling in one inertial frame could allow
backward in time signalling in another inertial frame; there
always exists some
(orthochronous) Lorentz transformation $L$ which transforms a space-like
wordline with $t>0$ into one with $t<0$.

Given backward in time signalling, a classical time paradox
can be
formulated: Assume two observers $A$ and $B$ which can communicate
{\em backward} in time;
 i.e., a signal traverses the
distance
$x_{AB}$ between them in time $t_{AB}$ such that
$t_{AB}<0$.
Then the observer $A$ might emit a signal at time $t_A$, which arrives
the observer
$B$ in $t_B$, where it is reflected and is back at the observer $A$
at a time $t_{A'}<t_A$, i.e., {\em before} observer $A$ has emitted the
original signal.
 If one performs a ``diagonalization,''
i.e., if one assumes that observer $A$ emits a signal at time
$t_A$  if and only if
{\em no} signal is absorbed at $t_{A'}$;
observer $A$ emits {\em no} signal at time
$t_A$ if and only if
a signal is absorbed at $t_{A'}$, one ends up with the simplest form of
time paradox.

The syntactic structure of this
paradox closely resembles Cantor's  diagonalization method
(based on the ancient liar paradox),
 which has been
applied by G\"odel, Turing and others for undecidability proofs in a
recursion theoretic setup.

How could the {\sc smorons} cope with such a time paradox?


One could argue that sound consists of elementary constituents (such as
atoms or molecules or clusters thereof),  whose motion is ultimately
governed by electromagnetic forces. Therefore, the valid theory is
electromagnetism, and any theory ``shell'' (``level'') such as the
{\sc smoron} theory of sound waves must ultimately be based  upon
(although
not necessarily be totally derivable from) it. As has been pointed out
earlier,
such a ``shell'' is very similar to what Anderson \cite{anderson} calls
``emerging law'' (cf. Schweber \cite{schweber}). In this
picture, the elementary constituents are a sort of ``hidden parameters''
for {\sc smoron} physics; something they can neither observe nor
control.
Therefore, the {\sc smoron} physics does not apply to configurations in
which
their physics ``shell'' is inappropriate. They are unable to control the
events. This is why the theory is cryptodeterministic, and the {\sc smorons}
have no free will with respect to diagonalization---they will simply not
be able to operationalize diagonalization purely in terms of sound
waves.

 One major goal of these considerations is the assertion that space and
 time are not God-given, metaphysical objects, but are subject to
 theoretical {\em construction}. The construction of space-time depends
 on {\em conventions}. Any inconsistency, for instance the possibility to
 construct time paradoxes, may be perceived as a problem of the improper,
 unfaithful construction of space and time rather than the impossibility
 to do certain tasks such as super-fast signalling. Any unfaithful
 construction of space-time may in turn be deeply rooted in the status
 quo of physical theory.

 \appendix
 \section*{Appendix A: Transformation rules of a CA light clock}
 \label{a:lightclock}

 $\varphi (>,\_,X) = >$, $\varphi (X,\_,<) = <$, $\varphi (\_,\_,\_) =
 \_$, $\varphi (X,\_,>) = \_$, $\varphi (<,\_,X) = \_$, $\varphi
 (\_,>,\_) = \_$, $\varphi (\_,<,\_) = \_$, $\varphi (\_,>,{\tt I}) = <$,
 $\varphi ({\tt I},<,\_) = >$, $\varphi (>,{\tt I},X) = *$, $\varphi
 (<,*,X) = {\tt I}$, $\varphi (X,<,*) = \_$, $\varphi (* ,1,X) = 2$,
 $\varphi (*,2,X) = 3$, $\varphi (*,3,X) = 4$, $\varphi (*,4,X ) = 5$,
$\varphi (*,5,X) = 6$, $\varphi (*,6,X) = 7$, $\varphi (*,7,X) = 8$,
$\varphi ( *,8,X) = 9$, $\varphi (*,9,X) = 0$, $\varphi (*,0,X) = 1$,
$\varphi (0,\_,X ) = \_$, $\varphi (1,\_,X) = \_$, $\varphi (2,\_,X) =
\_$, $\varphi (3,\_,X) = \_$, $\varphi ( 4,\_,X) = \_$, $\varphi
(5,\_,X) = \_$, $\varphi (6,\_,X) = \_$, $\varphi (7,\_,X ) = \_$,
$\varphi (8,\_,X) = \_$, $\varphi (9,\_,X) = \_$, $\varphi (X,*,0) = *$,
$\varphi (X ,*,1) = *$, $\varphi (X,*,2) = *$, $\varphi (X,*,3) = *$,
$\varphi (X,*,4 ) = *$, $\varphi (X,*,5) = *$, $\varphi (X,*,6) = *$,
$\varphi (X,*,7) = *$, $\varphi (X ,*,8) = *$, $\varphi (X,*,9) = *$,
$\varphi (X,1,X) = 1$, $\varphi (X,2,X) = 2$, $\varphi (X,3,X) = 3$,
$\varphi (X,4,X) = 4$, $\varphi (X,5,X) = 5$, $\varphi (X,6,X) = 6$,
$\varphi (X,7,X) = 7$, $\varphi (X,8,X) = 8$, $\varphi (X,9,X) = 9$,
$\varphi (X,0,X) = 0$, $\varphi (X,{\tt I},0) = {\tt I}$, $\varphi
(X,{\tt I},1) = {\tt I}$, $\varphi (X,{\tt I},2) = {\tt I}$, $\varphi
(X,{\tt I},3) = {\tt I}$, $\varphi (X,{\tt I},4) = {\tt I}$, $\varphi
(X,{\tt I},5) = {\tt I}$, $\varphi (X,{\tt I},6) = {\tt I}$, $\varphi
(X,{\tt I},7) = {\tt I}$, $\varphi (X ,{\tt I},8) = {\tt I}$, $\varphi
(X,{\tt I},9) = {\tt I}$, $\varphi (X,{\tt I},0) = {\tt I}$, $\varphi
(X,{\tt I},X) = {\tt I}$, $\varphi (* ,\_,X) = \_$, $\varphi (\_,\_,{\tt
I}) = \_$, $\varphi ({\tt I},\_,\_) = \_$, $\varphi ({\tt I},>,\_ ) =
\_$, $\varphi (\_,<,{\tt I}) = \_ \quad$.

$X$ stands for any state.

\end{document}